\documentclass[11pt]{article}
\usepackage{wds11,epsf,amsmath}

\lefthead{POLINOVSKYI \& MALYGIN}
\righthead{MEAN TRANSMITTED FLUX}

\begin{document}

\title{Mean Transmitted Flux in the Ly$\alpha$ Forest From a Sample of 2QZ Quasars}

\author{G. Polinovskyi}

\affil{Main Astronomical Observatory of National Academy of Sciences of Ukraine, Kiev, Ukraine}

\author{M. Malygin}

\affil{National Taras Shevchenko University of Kyiv, Physics Faculty, Kyiv, Ukraine}

\begin{abstract}

The power spectrum of transmitted flux in the Ly$\alpha$ forest in spectra of distant quasars gives us an information about matter power spectrum on the smallest spatial scales which is very important for testing of different models of dark matter. Only a few independent samples of spectra obtained on diferent instruments have been used for this purpose, thus using of additional independent sample is useful for verification an existing results. We used the data from the 2dF (Two-degree Field) QSO Redshift Survey (2QZ) to obtain the mean transmitted flux for three different redshift bins. After visual inspection and rejection of spectra with broad absorption lines, damped Ly$\alpha$ systems and low signal-to-noise ratio spectra our final sample contains 655 quasars with redshift range 2.3 $<z<$ 2.9. We present the results on composite spectra, determination of continuum level for different redshift ranges and the mean transmission in the Ly$\alpha$ forest as a function of redshift within the range 2.1 $<z_{forest}<$ 2.5. The obtained redshift-dependence of the mean transmission agrees well with the results of other authors. These results can be used for calculation of the flux power spectrum of Ly$\alpha$ forest.
\end{abstract}

\begin{article}

\section{Introduction}

The prominent absorption features of the spectra shortward of the Ly$\alpha$ emission line are called Ly$\alpha$ forest and were first discovered by Gunn and Peterson in 1965 like a small decrement in the spectra [\markcite{{\it Gunn et al.}, 1965}]. The high resolution measurements have been largely combined with hydrodynamical simulations and semi-analytical models. The results of these simulations have shown that the neutral gas responsible for the absorption is in a relatively low density, smooth environment, which implies a simple connection between the gas and the underlying dark matter. Thereby the fluctuations in the Ly$\alpha$ optical depth should reflect the density of the dark matter distribution on scales larger than a ``filtering'' scale related to the Jeans length. This paradigm for the origin of the Ly$\alpha$ forest has led to considerable interest in using quasar absorption spectra to study the dark matter distribution. Of particular interest is thereby the possibility to probe the density fluctuations of the matter with the flux power spectrum (FPS) of quasar absorption lines.

Only a few independent samples of spectra obtained on diferent instruments have been used for this purpose. In [\markcite{{\it Kim et al.}, 2004}] LUQAS sample consisting of 27 spectra obtained with the ESO Ultra-Violet Echelle Spectrograph (UVES) on VLT, Paranal, Chile, was used for calculations of the Ly$\alpha$ forest FPS for three redshift bins with $\langle z\rangle$~=~1.87, $\langle z\rangle$~=~2.18 and $\langle z\rangle$~=~2.58. This sample has spectral resolution $\sim$~45~000, signal-to-noise ratio larger than 25 and wavelength coverage 3000 -- 11000~\AA{}.
In [\markcite{{\it Viel et al.}, 2004}] the same sample and additional data from [\markcite{{\it Croft et al.}, 2002}] together with a large suite of high-resolution large box-size hydrodynamical simulations were used to estimate the mean transmitted flux and FPS for three redshift bins with $\langle z\rangle$ = 2.125, $\langle z\rangle$ = 2.44 and $\langle z\rangle$ = 2.72.
In [\markcite{{\it Cappetta et al.}, 2010}] the sample of 15 quasar spectra, observed with UVES was used to obtain the auto- and cross-correlation functions of the Ly$\alpha$ forest with median redshift $\langle z\rangle \approx$~1.8. This sample has signal-to-noise ratio $\sim$ 3 -- 12.
On the other hand, \markcite{{\it McDonald et al.}, [2006}] used 3035 quasar spectra from the Sloan Digital Sky Survey (SDSS), obtained with multi-fibre spectrograph that has resolution $\sim$~2000, wavelength coverage 3800 -- 9200~\AA{} and is placed at Apache Point Observatory, New Mexico, USA. The authors presented FPS of the Ly$\alpha$ forest for 9 redshift bins within the range of 2.2~$<z<$~3.8. SDSS spectra have much lower resolution and signal-to-noise ratio than UVES ones, but they cover much larger redshift range and count much more spectra that plays a significant role in the statistical errors reduction.

Thereby, it is useful to involve the independent data samples obtained on different equipment in the large-scale structure investigations with Ly$\alpha$ forest.

One of the main steps in procedure of Ly$\alpha$ forest FPS and correlation function calculations is estimation of the mean transmitted flux. In the present work we use the independent data from the 2dF (Two-degree Field) QSO Redshift Survey (2QZ) to obtain the mean transmitted flux for three different redshift ranges.

\section{The sample}

The 2dF facility used multi-fibre spectrograph on the 3.9-m Anglo-Australian Telescope, which is capable of observing 400 objects simultaneously over a 2$^\circ$ diameter field of view.
Two separate but linked large redshift surveys were the initial main projects with 2dF: one for 250,000 galaxies and one for 30,000 color-selected quasars covering redshifts up to 4.
Two spectographs provide spectra with resolutions of between 500 and 2000, over wavelength range of 4400 -- 11000~\AA{}. The main survey regions are two declination strips, one in the southern Galactic hemisphere spanning 80$^\circ \times $15$^\circ$ around the South Galactic Pole, and the other in the northern Galactic hemisphere spanning 75$^\circ \times $ 10$^\circ$ along the celestial equator; in addition, there are 99 fields spread over the southern Galactic cap. The survey covers 2000 deg$^2$ [\markcite{{\it Lewis et al.}, 2002}].

Our initial sample consists of 1,961 non flux-calibrated spectra and covers the redshift range of 2.3 $ < z < $ 3.3. For calibration of the spectra we used the 2dF relative efficiency curve from [\markcite{{\it Lewis et al.}, 2002}]. Calibration included the average telluric lines and fibre absorption. Spectra with low signal-to-noise ratio, non-quasar spectra, spectra with broad absorption line features (BAL quasars) and the damped Ly$\alpha$ systems (DLAs) were removed from the initial sample. The DLAs are the systems with a number density of $>$2$\times10^{20}$ cm$^{-2}$ that is three orders of magnitude higher than for the Ly$\alpha$ forest systems [\markcite{{\it Meiksin et al.}, 2009}] and appears in the spectrum as a wide and deep absorption feature in the Ly$\alpha$ forest region. Thus such spectra with DLAs can affect the value of the mean transmission. BALs from material along the line of sight to the quasar are generated by high-velocity, structured outflows in the central regions of quasars. Absorption signatures of these outflows are, by definition, at least 2000~km~s$^{-1}$ wide and are blueshifted [\markcite{{\it Hall et al.}, 2002}], [\markcite{{\it Gibson et al.}, 2010}]. The spectra containing BALs can heavily obscure the continuum in the resulting composite spectum. After calibration and cleaning the resulting sample contains 655 spectra with redshifts within the range of 2.3~$ < z < $~2.9.

\section{The Method}

\subsection{Generating the composite spectra}

Subtle global spectral properties and features can be studied by combining large numbers of quasar spectra into composites. Such composite spectra have a high signal-to-noise ratio and reveal weak emission features that are rarely detectable in individual spectra. This method is used by many authors for both high and low signal-to-noice data sets (see [\markcite{{\it Telfer et al.}, 1997}], [\markcite{{\it Brotherton et al.}, 2001}], [\markcite{{\it Scott et al.}, 2004}], [\markcite{{\it Francis et al.}, 1991}] and others). We have made composite spectra to obtain the value of the mean transmitted flux for each of them.

Before making composites all spectra were smoothed with 3-point unweighted sliding-average smooth, rebinned to the rest frame and flux normalized. Normalization procedure involved the next steps: estimation of the mean flux within the range of 1450~\AA{}~$<\lambda_{rest}<$~1470~\AA{} for each spectrum and division each spectrum by this value. This spectral range was chosen as the most free one from emission lines.

We have compiled three composite spectra as arithmetic means for redshift ranges of 2.3~$< z <$~2.5, 2.5~$< z <$~2.7 and 2.7~$< z <$~2.9. The resulting composite spectra cover the rest frame wavelength range of 1056 -- 1855~\AA{}. It is worth to note that the arithmetic mean of the power-law spectra will not, in general, result in a power law with a mean exponent but \markcite{{\it Vanden Berk et al.}, [2001}] has shown that the difference in the spectral exponent for mean arithmetic and mean geometric composite spectra is insignificant. Such a difference is not important for our purpose.

\subsection{Continuum fitting}

The estimation of the continuum level is a very delicate step in the process of spectra reduction. There is no specific and adopted method for continuum fitting procedure in the literature, although different methods had been used (see [\markcite{{\it Press et al.}, 1993}], [\markcite{{\it Cappetta et al.}, 1993}], [\markcite{{\it Vanden Berk et al.}, 2001}] and others). The problem of fitting the quasar continuum is complicated by the fact that there are essentially no absorption-line-free regions in the spectrum. We have determined the continuum level for whole composite spectrum using a set of regions which are free from emission features (judged by eye). The regions that satisfy this are 1281 -- 1291~\AA{}, 1321 -- 1327~\AA{}, 1348 -- 1361~\AA{} and 1438 -- 1468~\AA{}. Continuum level was fitted with a power-law $f = A\lambda^{\alpha}$. The composite spectrum for 2.7~$<z<$~2.9 with determined continuum is shown in Figure~1. The uncertainties for the normalizing coefficient A and for the slope $\alpha$ can affect the results reported in this paper on the mean transmitted flux function and should be considered as a source of possible systematic errors in addition to quoted statistical errors of the results.

\begin{figure}[htb]
\begin{center}
\begin{tabular}{c}
  \epsfxsize=82mm
  \epsfbox{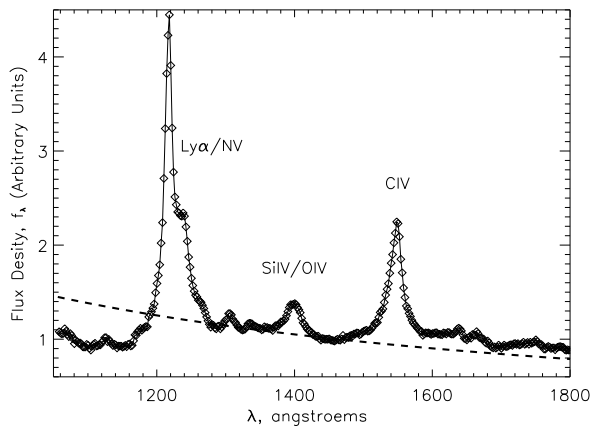}
\end{tabular}
\end{center}
\caption{The composite spectrum for the 2.7~$< z <$~2.9 range (solid line with data points) and estimated continuum (dashed line).}
\end{figure}

\subsection{The mean transmitted flux}

The matter density fluctuations can be calculated from the FPS of Ly$\alpha$ forest which in turn is obtained using the mean transmitted flux in this region of spectrum. The flux power spectum is the measure of the variance in the amplitude of the Fourier transform coefficients of the transmitted flux.

The transmitted flux of a quasar spectrum is given by $F'(\lambda)=F_{cont}(\lambda)e^{-\tau(\lambda)}$, where $F_{cont}(\lambda)$ is the continuum, $e^{-\tau}$ is the transmission and $\tau(\lambda)$ is the opacity or optical depth. For the spectrum reduced by the continuum fitting:
\begin{equation}\label{transm_flux_red}
 F(\lambda)=e^{-\tau(\lambda)}+n(\lambda),
\end{equation}
where $n(\lambda)$ describes the stochastic noise. Averaging (\ref{transm_flux_red}) over the density fluctuations (over all pixels in a redshift bin) one can obtain $\bar{F}(\lambda)=\left\langle e^{-\tau(\lambda)}\right\rangle$. Generally speaking, this value is position dependent, as it depends on the photoionization rate of HI and the temperature of the intergalactic medium. If one assumes that $\bar{F}(\lambda)$ is a constant, this value then is the mean transmitted flux $\bar{F}(z)$ which is a function of the redshift.

The mean transmitted flux can be determined through the folowing considerations. The flux density in the $i$-th pixel in a single spectrum within the forest range is

\begin{equation}\label{mean_tr_flux_one}
  f_i^j(\lambda_{rest},z) = A^jC(\lambda_{rest})[\bar{F}(z) + \delta F_i^j] + n_i^j,
\end{equation}
where $j$ stands for spectrum, $A^j$ is the normalizing coefficient, $C(\lambda_{rest})$ is the continuum, $\bar{F}(z)$ is the mean transmitted flux, $\delta F_i^j$ is the fluctuations of the mean transmission and $n_i^j$ is the noise. Averaging over a small redshift bin (a set of spectra) for the composite spectra we obtain

\begin{equation}\label{mean_tr_flux_comp}
  f_i = \bar{C}(\lambda_{rest})\bar{F}(z),
\end{equation}
as $\langle\delta F_i\rangle\equiv0$ over all lines of sight and $\langle n_i\rangle\equiv0$ over all spectra. We determine the continuum level inside the forest region (1056 -- 1202~\AA{}) as

 \begin{multline}\label{Continuum}
  C=A \left(\lambda_{rest}\right)^{\alpha}+c_{1}\exp\left[-\frac{(\lambda_{rest}-1066.66)^{2}}{2c_{2}^{2}}\right]+c_{3}\exp\left[-\frac{(\lambda_{rest}-1123.00)^{2}}{2c_{4}^{2}}\right]+ \\ + c_{5}\exp\left[-\frac{(\lambda_{rest}-1175.70)^{2}}{2c_{6}^{2}}\right]+ c_{7}\exp\left[-\frac{(\lambda_{rest}-1215.67)^{2}}{2c_{8}^{2}}\right],
 \end{multline}
where A and $\alpha$ are the mean continuum coefficients, the wavelenghts of 1066.66~\AA{}, 1123.00~\AA{}, 1175.70~\AA{} and 1215.67~\AA{} are the wavelengths of ArI, FeIII:, CIII$^*$ and Ly$\alpha$ spectral lines respectively.


\section{Results}

For three redshift bins 2.3~$<z<$~2.5, 2.5~$<z<$~2.7 and 2.7~$<z<$~2.9 we derive continuum slopes of $\alpha\approx$--0.6, $\alpha\approx$--0.5 and $\alpha\approx$--1.1 respectively.
The resulting mean transmitted flux and parameters of the lines are presented in Table 1. Estimated 1$\sigma$ error for $\bar{F}$ includes statistical errors of the fitted data. For calculation of the mean transmitted flux mean continuum parameters were adopted. For illustration in Figure~2 the Ly$\alpha$ forest region with the obtained fitting curve are shown for 2.5 -- 2.7 redshift bin.

\begin{table}[t]
\scriptsize
\centering
\caption{Line parameters used for continuum fitting and the mean transmitted flux for three redshift bins.}
\centering
\begin{tabular}{|c|c|c|c|c|c|c|c|c|c|c|}
\hline
Quasars redshift & $z_{forest}$ & \multicolumn{2}{c|}{ArI} & \multicolumn{2}{c|}{FeIII:} & \multicolumn{2}{c|}{CIII$^*$} & \multicolumn{2}{c|}{Ly$\alpha$} & $\bar{F}$ \\
\cline{3-10}
bin & & c1 & c2 & c3 & c4 & c5 & c6 & c7 & c8 &\\ [3 pt]
\hline
2.3-2.5 & 2.13 $\pm$ 0.29 & --- & --- & 0.2 & 30.5 & 0.1 & 11.3 & 1.5 & 15.2 & 0.83 $\pm$ 0.04 \\
\hline
2.5-2.7 & 2.29 $\pm$ 0.30 & 0.3 & 22.6 & 0.04 & 6.8 & 0.2 & 9.2 & 1.7 & 16.2 & 0.78 $\pm$ 0.03 \\
\hline
2.7-2.9 & 2.47 $\pm$ 0.31 & 0.2 & 10.5 & 0.1 & 6.4 & 0.2 & 7.0 & 1.5 & 16.2 & 0.69 $\pm$ 0.03\\
\hline
\end{tabular}
\end{table}
\begin{figure}[!h]
\begin{center}
\begin{tabular}{c}
  \epsfxsize=82mm
  \epsfbox{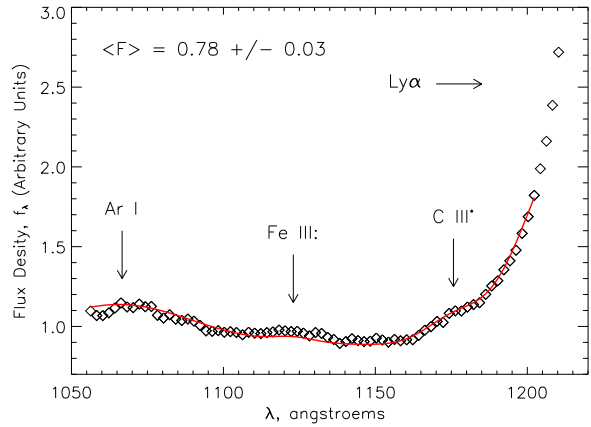}
\end{tabular}
\end{center}
\caption{Ly$\alpha$ forest region (data points) with spectral lines (pointed with arrows) and obtained fitting curve (solid line) for composite spectrum of the 2.5~$<z<$~2.7 redshift bin.}
\end{figure}
\begin{figure}[!h]
\begin{center}
\begin{tabular}{c}
  \epsfxsize=82mm
  \epsfbox{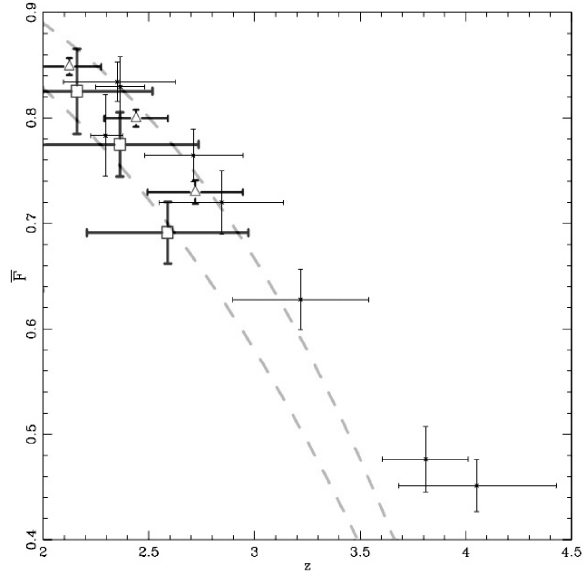}
\end{tabular}
\end{center}
\caption{Obtained mean transmitted flux compared with other authors results (for the explanation see the text).}
\end{figure}

\vspace{7pt}

In Figure 3 we present the comparison of the obtained mean transmitted flux with the results of other authors. Results of \markcite{{\it McDonald et al.}, [2000}], where eight high-resolution and high signal-to-noise ratio spectra obtained with HIRES instrument were used, are shown with dots. Results of \markcite{{\it Viel et al.}, [2004}] where the mean transmitted flux was obtained for three redshift bins using the spectra from the LUQAS sample, are shown with triangles. Two dashed lines show the upper and lower limits for $\gamma$ in the power-law describing the mean transmitted flux evolution described by \markcite{{\it Press et al.}, [1993}] as function $\bar{\tau}(z) = A(1+z)^{\gamma+1}$, where the coefficients $\gamma = $  2.46 $\pm$ 0.37 and $A = $ 0.0175 -- 0.0056$\gamma$ $\pm$ 0.0002 were obtained using this power-law to fit the data. Authors used 29 spectra from the Schneider-Schmidt-Gunn (SSG) sample. Our results are shown with squares. The error bars in $z$ direction show the Ly$\alpha$ forest range in each bin.

\section{Conclusions}

Independent result of calculation of the mean transmitted flux was obtained for three redshift bins using the low S/N and low resolution data from the 2dF QSO survey. Mean transmitted flux is $\bar{F}(z) = $ 0.83~$\pm$~0.04, $\bar{F}(z) = $ 0.78~$\pm$~0.03, $\bar{F}(z) = $ 0.69~$\pm$~0.03 for the redshift bins of the Ly$\alpha$ forest $\bar{z}_{forest}$~=~2.13, 2.29 and 2.47 respectively. The obtained redshift-dependence of the mean transmission agrees well with the results of other authors (e.g. \markcite{{\it McDonald et al.}, [2000}], \markcite{{\it Viel et al.}, [2004}], \markcite{{\it Press et al.}, [1993}]). It is notable that the derived results on the mean transmission have lower values in comparison with other authors. This can be explained by the uncertainties of the continuum level estimation and usage of the simple power-law for this purpose. Besides, we have shown that the data from 2QZ is reasonable for the Ly$\alpha$ forest investigations. These results can be used for calculation of the flux power spectrum of Ly$\alpha$ forest.


\end{article}

\end{document}